\documentstyle[seceq]{ptptex}

\title{Observational Tests of the Topology of Planckian Space-Time}

\author{Marco {\sc Spaans}\footnote{Hubble Fellow, E-mail:
mspaans@cfa.harvard.edu}}

\inst{Harvard-Smithsonian Center for Astrophysics, 60 Garden
Street, MS 51, Cambridge, MA 02138, USA}

\recdate{
\today
}

\abst{
Observational diagnostics are constructed which reflect the underlying
topology of Planckian space-time, and are directly related to phenomena on much
larger scales. Specific predictions are made for the masses of elementary
particles, coupling constants, quark confinement, black hole states,
and the cosmological constant. In particular, a mass of 131.6 GeV is found
for the Higgs boson.\hfill\break\indent
The main presented result is a discrete spectral signature at inverse
integer multiples of the zero point frequency $\nu_0=857.3588$ MHz
(34.96698 cm).
That is, each photon of frequency $\nu_0m$, for an integer $m$, is paired with
an otherwise identical photon $\nu_0/m$, produced by the vacuum, but not vice
versa. The reader interested in the latter result only, should proceed to
Section 6 after the Introduction.
}

\begin{document}

\maketitle

\section{Introduction}

Recently, it has been proposed that space-time on the Planck scale exhibits a
definite non-trivial topology generated by the three-torus $T^3$ and the handle
$S^1\times S^2$ [1,2]. The interested reader is referred to these papers for an
exhaustive description of the resulting theory. Only those elements relevant
to the investigations of this work will be discussed in more detail in the
subsequent sections. A summary, not aimed at being comprehensive,
of the formalism in [1,2] is as follows. A thought
experiment leads to the ground state of the Universe as the Lorentz invariant
topological manifold $Q=2T^3\oplus 3S^1\times S^2\equiv P\oplus\Theta$ with a
$\lambda\phi^4+\mu^2\phi^2$ self-interaction potential, and an equation of
motion
$$4\pi i\partial_\nu\phi [(\partial^\nu q_\lambda )q^\mu q_\mu -(\partial^\nu
q_\mu )q^\mu q_\lambda ]=q_\lambda q^\mu\Box q_\mu -q^\mu q_\mu \Box
q_\lambda ,\eqno(1a)$$
$$q^\mu q_\mu ={\rm cst},\eqno(1b)$$
for the four component wave function $q_\lambda$.
The manifold $Q$ has $23=2\cdot 7+3\cdot 3$ (14 for $P$, 9 for $\Theta$)
degrees of freedom, corresponding to the number of
elementary particles which depend on the homotopy of space-time, i.e.\
excluding the photon and graviton. The intrinsic measures of $Q$, $P$ and
$\Theta$ are $23^{-3}$, $14^{-3}$ and $7^{-3}$, respectively, where the cubic
powers reflect the three-fold identifications of the three-torus.
When the mini black holes evaporate the {\it lattice} of three-tori, $L(T^3)$,
is described by the equation of motion
$$q^\mu q_\mu\Box q_\lambda =q_\lambda q^\mu\Box q_\mu ,\eqno(2)$$
until the temperature of the Universe drops below the grand unification energy.
The wave function $q_\lambda$ has four components
corresponding to the $e_{\rm M}=4$ homotopically inequivalent paths through
$L(T^3)$. The periodic structure of $L(T^3)$ implies
$$q_\lambda (x,y,z,t)=q_\lambda (x+L_1,y+L_2,z+L_3,t),\eqno(3)$$
as boundary conditions, i.e.\ {\it topological identifications}, for the wave
function, with $L_i=n_i\ell_{\rm Planck}$ and $n_i$ positive integers. It is
the structure $L(T^3)$ and the boundary conditions (3) which lie at the heart
of (Planck scale) quantum physics.

In [1,2] predictions were made unique to this topological approach, indicated
as topological dynamics (TD) from hereon, and those already observable were
found to be consistent with current observational limits. A summary of the
predictions needed in this work is
1) The unified theory generated by $Q$ and $L(T^3)$
possesses a $SU(5)$ gauge group, $Z_3$
and $Z_2\times Z_3$ discrete groups, and an anomalous $U(1)$ symmetry.
2) The cosmological constant (on the left hand side in the Einstein equation)
is equal to 2$m_{\rm Planck}$ times the number density of (macroscopic) black
holes in Planck units, and is therefore {\it almost} zero. 3) The electron
mass equals $1+{\cal A}/3m^0_{\rm e}$ with ${\cal A}=(14/23)^3$ and
$m^0_{\rm e}=m_{\rm Planck}/23!$ up to {\it position dependent} quantum
corrections discussed below, and the mass of the electron neutrino is 14! times
smaller than that of the electron. 4) The GUT energy scale is
$m_{\rm Planck}/138$, $K=1/138$ in Planck units, and partitioned over the 7
degrees of freedom of the three-torus. 5) The typical amplitude of the initial
mass-energy fluctuations at the end of the GUT phase is $23^{-3}$.
6) The numerical factor of $1/4$ in the expression of the black hole entropy is
fixed by the topology of $L(T^3)$ and equals $e_{\rm M}^{-1}$. 7) The masses of
elementary particles can vary with position at a typical amplitude of $7^{-3}$
on cosmological scales, according to the probability distribution
$q^\mu q_\mu$. 8) CP violation is mediated through time reversal violating
vibrations in $L(T^3)$ with an energy $E_0=3.546\times 10^{-6}$ eV.\hfill\break
The present work is
concerned with the derivation of additional results in the context of TD.
Since TD only has the speed of light, Planck's constant, and the gravitational
constant as input parameters, {\it any} disagreement with observations or
experiment will rule it out.

This paper is organized as follows. Section 2 gives
predictions for the masses of elementary particles.
Section 3 extends the notion of [2] that the inertia of
material particles is a position dependent property of space-time itself
through $q^\mu q_\mu$, and
discusses some cosmological ramifications. Section 4 presents an exposition of
the quantum states intrinsic to black holes. Section 5
discusses the spin lattice structure of space-time and
its implications for QCD. Section 6 suggest a direct
measurement of the physical reality of $Q$ and $L(T^3)$ through the dual
partners of photons. Section 7 contains the conclusions.

Sections 2 through 5 are essentially (sometimes tedious)
enumerations of the degrees of
freedom of $Q$ and $L(T^3)$. The reader interested in the spectral signature,
a feature clearly outside of the Standard Model, therefore should go to
Section 6 directly.

\section{The Masses of Elementary Particles}

The $SU(3)$ gauge potential of QCD is realized
on the junctions of the lattice[2], and therefore acts on pairs of quarks.
Since each three-torus in $L(T^3)$ has six neighbours, there are $6!$
physically distinct orderings for the adjacent quarks. It follows that the
doublet (u and d are both charged) mass ground state is\hfill\break
$m_{\rm u,d}=6!m^0_{\rm e}=340$ MeV. The s, c, b, and t excited mass
states represent four topologically distinct configurations of the lattice,
which yield mass multiplication factors given by the ratios of degeneracy
factors $\delta_{\rm q}/\delta_{\rm u,d}$, with $\delta_{\rm u,d} =2$. One
finds $\delta_{\rm s}=3$ and $\delta_{\rm c}=2\delta_{\rm s}$ because there
are three fermionic states supported by any three-torus[2], and twice that many
by the doublet. It further follows that $\delta_{\rm b}=\delta_{\rm s}^3$ due
to their possible combinations into triplets. Finally,
$\delta_{\rm t}=4/3\delta_{\rm s}^6$ for the doublet, where there
are 4 homotopically inequivalent paths between the two three-tori, but 3 of
these are associated with $T^3$ itself (the 1 is of course $S^3$). From these
numbers one finds rest masses for s, c, b, and t given by\hfill\break
$m_{\rm s}=510$ MeV,
$m_{\rm c}=1.02$ GeV, $m_{\rm b}=4.59$ GeV, and $m_{\rm t}=165.3$ GeV
respectively, in good agreement with available experiments.

The excited lepton states scale with the inverse of the anomalous
$U(1)$ symmetry breaking scale[2], $\beta^{-1}=9K=9/138$, in Planck units.
The ratios of the excited state masses to the ground state must be double
and triple powers of $\beta$ because the equation of motion (2) supports cubic
self-interactions, with a quadratic scalar term. This
yields masses for the muon and tau of 119.4 MeV and 1831 MeV, respectively,
up to topological mass corrections, in good, 13\% and 3\%, agreement with
observations. For respective their neutrinos one finds $1.37\times 10^{-3}$ eV
and $0.021$ eV. These results are consistent with SuperKamiokande findings[3].

Analogous to [2], there is a purely topological correction to these
excited state masses due to the structure of $Q$ and $L(T^3)$. The transition
$Q\rightarrow T^3$, where $Q$ contains two three-tori, yields an intrinsic
accuracy ${\cal B}=(7/14)^3$. This number
signifies the information regarding the $\mu$ state which needs to be recovered
after the transition to the excited state.
The sign of this mass correction is thus always negative,
i.e.\ $-{\cal B}m_\mu$.
One finds $1/4$ this number for the $\tau$ because the latter
is a third order state, and can have any of the four paths through
$L(T^3)$ attached to a (quadratic) $S^1$ loop. The same corrections
apply to the associated neutrinos, but not to the quarks since they are not
part of the electroweak sector. Hence the use of $m^0_{\rm e}$ in the doublet
quark ground state above.
Once applied, these corrections yield approximately $\sim 7^{-3}$ agreement
with experimental results,\hfill\break
$m_\mu =104.5$ Mev, $m_\tau =1774$ MeV, $m_{\nu_\mu}=1.20\times 10^{-3}$ eV,
$m_{\nu_\tau}=0.02$ eV.
In Section 5 the electroweak mixing angle and fine
structure constant are computed, and the topological masses of the vector
bosons are presented there.

Furthermore, the mass of the Higgs
boson follows from $\lambda^{-1}=7$, the number of degrees of freedom of an
individual three-torus[2], and $m_{\rm H}=\surd (2v^2\lambda )$ with the Fermi
coupling constant $G=1.16632\times 10^{-5}$ GeV$^{-2}=v^{-2}/\surd 2$ (see
Section 5). This yields a value for the Higgs boson mass of\hfill\break
$m_{\rm H}=131.6$ GeV, consistent with current experimental limits.

Note that the probability distribution $q^\mu q_\mu$ for the rest masses of
elementary particles is the same for all particle species, but that collapse
proceeds independently for the ground and excited states because they are
associated with topologically distinct configurations on $Q$ and $L(T^3)$.
Given that our Galactic environment is the result of a primordial density peak,
it is to be expected that most of the expectation values computed above are
increased by a few times $7^{-3}$.

\section{Spatial Variations in Inertia}

\subsection{Non-Gaussianity}

The equation of motion (2) on $L(T^3)$ is solved by solutions to the linear
Klein-Gordon equations
$$\Box q_\lambda =-m^2q_\lambda,\eqno(4)$$
where the solution space is restricted by (3), and (4) can be rewritten as a
general Proca equation for $m^2\not= 0$ since the latter then yields
$\partial^\mu q_\mu$. The solutions to (4) are
taken here to be quadruple travelling wave forms for $m^2=0$
$$q_\lambda =f_\lambda (l_1x+l_2y+l_3z-t)+f_\lambda (l_1x+l_2y+l_3z+t),
\eqno(5)$$
where the direction cosines $l_i$ satisfy
$$l_1^2+l_2^2+l_3^2=1,\eqno(6a)$$
the functions $f_\lambda$ have the standard forms for this wave
equation limit, and the boundary conditions (3) further require that
$$l^iL_i=g^{ij}l_jL_i=0,\eqno(6b)$$
with $g^{ij}$ the three-metric which relates the topological boundary
conditions to the geometry of the Universe. The
coefficients $L_i$ determine the physical shape of the wave forms.
It is the superposition $S$ of this set of solutions over all possible $L_i$,
and the $l_i$ they allow, which determines the inertia of material particles
in the Universe. Note that the functional forms in normalized coordinates of
the $f_\lambda$ generally will be different, depending on the solutions to (1).

The superposition of the $q_\lambda (l_i[L_i])$ through the dependent
coefficients $l_i$ need not be Gaussian. In fact, an investigation of the
constraints for $g^{ij}=A(t)\delta^{ij}$ with a scalar $A$ shows that $S$ has
non-Gaussian features because very anisotropic wave packages (lines and sheets)
are part of the solution space, and move almost orthogonally to their major
axis/axes, {\it if} one assumes that there are no preferred identifications
between various three-tori. As all these sheets and lines are
superposed, one creates small scale and large amplitude fluctuations in both
the mass-energy {\it and} particle rest masses.

Note that a period of inflation is straightforwardly incorporated through the
relation (6b) since it is invariant under a constant scale transformation.
The possible durations $\Delta t$ over which inflation occurs are determined
by solutions to (1a) and (1b) which satisfy $\partial_\nu \phi =0$ at some
time $t$, and fix the time interval $\Delta t=t-t'$ associated with
$|\phi_t-\phi_{t'}|=1/3$ (see [2]).

\subsection{Primordial Fluctuations}

Observe with respect to the different amplitudes for mass-energy and
rest mass fluctuations the interesting transition at decoupling, when the
photon energy density no longer dominates that of the matter. The respective
amplitudes for mass-energy and rest mass perturbations, are $23^{-3}$ and
$7^{-3}$[2]. These differ by a factor of approximately 35. To satisfy the
$23^{-3}$ constraint, one thus requires spatial variations in the number
density of massive particles on the order of $7^{-3}$. At the epoch of
decoupling these number density fluctuations induce particle diffusion
from number overdense to number underdense regions, and thus selectively
increase (bias) the amplitudes of mass overdense regions.

\section{Black Holes}

For black holes the solutions to (4) are taken to be generic spherical waves
$$q'_\lambda =\Sigma_{k,s}(2\omega_k V)^{-1/2}\epsilon_{\lambda s}(k)[a_s(k){\rm e}^{-ikr+
\omega_k t}+b^\dagger_r{\rm e}^{ikr-\omega_k t}],\quad s=1,2,3,\eqno(7)$$
with orthonormal polarization vectors $\epsilon_s$, the Lorentz condition,
wave vector $k$, proper volume
$V$, $\omega_k^2=m^2+k^2$, the standard interpretation in terms of annihilation
and creation operators, analogous expressions for photons with $m^2=0$, and the
index $\lambda$ for the four homotopically distinct paths on $L(T^3)$. For
the $S^1\times S^2$ topology of the handle one finds the boundary conditions
$$q'_\lambda (r=0,t)=q'_\lambda (r=a',t),\eqno(8)$$
in spherical coordinates, with $a'$ the Schwarzschild radius in Planck units.
Therefore, all the mass-energy crashing into the singularity is represented on
the event horizon. Note that the localization of the black hole's information
like this can only be established through the possible topological
identifications in $L(T^3)$, and is in fact required by the existence
of event horizons of arbitrary size which Nature must accommodate[2].
Furthermore, one has the parameterization of the wave vectors
$$k={{2\pi}\over{a'}}n, \quad n=\pm 1,\pm 2,...,\pm a',\eqno(9)$$
along the $S^1$ of the handle.

\subsection{The Cosmological Constant}

In [1] an algebraic derivation was given for the small value of the
cosmological constant $\Lambda$. The central result was that $\Lambda$, viewed
as the spontaneous creation of pairs of mini black holes, is
proportional to the present number of macroscopic black holes. In [2] a thought
experiment was presented to incorporate this finding in a generalization of
Mach's principle. Note here the difference between the left and the right hand
side of the Einstein equation. Section 6 will deal with the right, the zero
point energy associated with the particle vacuum, and this section with the
left, curvature associated with a non-trivial, but massive, topology (which is
not fixed by general relativity). Clearly, this distinction must be arbitrary
at some fundamental level, as the following discussion and Section 6 will
examplify.

The boundary conditions are half the solution to the equations of motion. The
key to understanding $\Lambda$ is precisely in these, as follows.
The solutions (7) must match the rest of the Universe outside of the event
horizon if evaporation is to occur on thermodynamic grounds at some point in
time, i.e.\ (8) only describes the black hole as a closed system. Let the
collapse be denoted by $\{a'_i\}$ with $i=1$ corresponding to $r=0$. The length
of any body as a function of proper time in Planck units
$a''_i\propto(\tau [a'_i]-\tau [a'_1])^{-1/3}$ reflects the matching condition
for the surface $\tau [a'_i]$ which characterizes a black hole configuration
$i$. Because $V\propto a_i''^{-3}$ in (7), one finds a peak, the singularity,
in $q^\mu q_\mu$ for particles represented by (7) near $i=1$. Note that the
mass-energy {\it and} particle rest masses follow this peak. It is therefore
possible, as mentioned above, to concentrate all the information on the black
hole horizon through (8), while using the singularity as the locus of the
mass-energy. Since the resulting probability distributions are normalized, only
the $i=1$ configuration gives a non-zero contribution under
$\lim_{V\to 0}\int_V(q^\mu q_\mu )_i$.

Furthermore, $a''_i$, like $a'$, is a geometric quantity which incorporates the
shape of space-time. It is the function $a''_i$, similar to (6b), that
provides the relationship between the topology and the geometry of space-time.
That is, the information is contained in the wave modes (7), whose
absolute extend, like the scale invariance of (6b), derives from space-time
geometry. It follows that the modes associated with the black hole singularity,
$i=1$, extend over arbitrary distances. Clearly, such a conclusion again
depends crucially on the existence of the topological identification (8) on
$L(T^3)$ as discussed above. In conjunction, the rest masses diverge at $r=0$.
This allows Planck mass excitations to be generated throughout the Universe,
and render the singularity finite by associating it with any point beyond the
horizon, for a Planck time, while preserving the handle topology.
These are precisely the mini black hole pairs (both signs of $n$) which
constitute the left hand side cosmological constant.

The singularity thus realizes the formation and evaporation of the black hole
as a polizarization of the vacuum by the matter degrees of freedom. Although
this result might appear counterintuitive, the introduction of a topological
origin for the inertia of massive particles leads directly to 't Hooft's
suggestion that black holes are a natural extension of elementary particles,
and hence to a link between topology and the properties of the vacuum. When the
black hole has evaporated the singularity vanishes into that same vacuum.

\subsection{Information and States}

Note that the solutions (7) also solve the Dirac equation. In fact, the four
components of $q_\lambda$, a topological property, facilitate a perfect match
for the four Dirac spinor components, given the $SL(2;C)$ spin structure on the
four-manifold bounded by the nuclear prime manifold $T^3$.
Thus, the mass, charge, and spin of any elementary particle can be encoded in
the wave modes, and black hole
evaporation can proceed unitarily in this sense. Nevertheless, one will
need the complete evaporation history for the reconstruction of the data
\footnote{Suggestions by M.\ Bremer on these matters are greatfully
acknowledged.}.

It is straightforward to predict the number of black hole states $N_{\rm BH}$.
There are $2i$ wave modes (quanta) at a given $a'_i$, and 1 in 4 of these
contains relevant information since $\lambda =1..4$ allows the construction of
every double, triple and quadruple wave number of a mode through
multiplication. The state space of a black hole $a'$ then consists of all
configurations formed by the sum of its $a'_i$. This yields for an integration
over $4\pi$ steradians in Planck units
$$^7{\rm log}N_{\rm BH}={{1}\over{4}}\Sigma^{a'}_{i=1}8\pi i=
4\pi (a'^2+a')/4,\eqno(10)$$
with 7 the natural base number for $L(T^3)$, i.e.\ the smallest unit of
information is $ln 7$. For large $a'$ one has the semi-classical result for the
entropy $S_{\rm BH}=A/4=\pi (a'^2+a^2)$, with $A$ the area and $a$ the angular
momentum per unit mass. A comparison shows that only that part of the surface
area associated with the event horizon, $a'^2=[M+\surd (M^2-Q^2-a^2)]^2$ for a
charge $Q$, contributes to the number of quantum states (10). Therefore, the
ergosphere constitutes a region for interaction with the internal states (which
are quadrupally degenerate), that adds to the total entropy but whose
information content is intrinsically macroscopic. That is, $\pi a^2$ is
effectively an integration constant.

The black hole temperature $T$ associated with $A$ characterizes the
macroscopic entropy of the black hole system through the irreducible mass
$M^{\rm ir}=(A/16\pi)^{1/2}$. This temperature determines
the internal structure of the black hole, and is relevant to its statistical
properties on the lattice of three-tori. That is, the black hole emits a
spectrum constrained by the {\it summed}, because of (8), incoming wave
function, where the temperature $T$ determines the probability for re-emission
through ${\rm exp}[-E/kT]$. The energy $E$ is the particle's energy when it
originally entered the black hole.

\section{The Lattice Structure of Space-Time}

\subsection{Holonomy}

Due to the seven-fold multiplicity of the three-torus[2], one should view every
heptaplet of three-tori, {\it not} necessarily a Planck length apart, as a
concrete entity. From this, there is a direct correspondence, in a topological
sense, between the
$T^3$ heptaplet and the well known compactification space $K3\times T^3$. That
is, the heptaplet can also be interpreted as a (7+)3+1 space-time, where the
internal 7 must be of the form $M^4\times T^3$, with $M^4$ simply connected.
It must have a $H=SU(2)$ (rather than the generic $O(4)=SU(2)\times SU(2)$)
holonomy. This is because of the $U(1)$ factors on the junctions of $L(T^3)$
along any closed curve. Therefore one seems to find a compactification space
for 11-dimensional ${\cal N=1}$ supergravity. Nevertheless, when one considers
the structure of the heptaplet in full, the possible holonomies (see also
QCD below) reflect just the possible particle interactions.

\subsection{QCD}

Recall that the spin structure of the nucleon
uncovered by recent experiments suggests that the three-quark model could be
incomplete. This might imply that all six quarks are merely individual
manifestations of what is intrinsically a global background structure. In fact,
it is easy to see that the $U(1)\sim O(2)$ rotation group on the
{\it junctions} of the cubic $L(T^3)$ lattice, leads to {\it sub}-structures
given by two-dimensional periodic Ising models. This is not surprising since
$SU(3)\supset U(1)^3$. One also finds that this system as a whole is
{\it frustrated}\footnote{The author is very greatful to B.\ Canals for
discussions on this point.}, which is a fascinating realization given the
importance of the strong interaction in the Universe.

The very presence of frustration {\it requires} the existence of collective
excitations on the Ising driven lattice. This provides a physical basis for the
concept of confinement under the QCD gauge group.
The $SU(3)$ gauge symmetry yields a nearest neighbor Hamiltonian,
whose extrema need to be determined, given by
$$h=\kappa\Sigma_{i,j;i\not= j} [S_i^kS'_{ik}][S_j^kS'_{jk}],\eqno(11)$$
where $S_k=S'_k$ is excluded if it holds for all $k$ up to a global sign,
$k=1..3$ labels the three twist (``color'') variables
$S_k=\pm 1$, and periodic boundary conditions need to be applied if the
Universe is closed. The form (11) is invariant under transformations
$S_k''=MS_k$
with $M\in SU(3)$, i.e.\ unlike a photon a gluon carries a (color) charge and a
quark can therefore interact with the gluons it emits.

The 8 gluons are a direct consequence of the number of different $S_k$.
It is easy to see that there are eight triplets $S_k$. The contraction of
various $S_k$ yields, when grouped together in doublets differing only by an
overall sign, 12 possibilities, i.e.\ 6 quarks with their anti-particles.
In the end, $S^kS'_k=\pm 1$ yielding a true spin lattice. Of course, due to
the $SU(3)$ gauge group and the conditions on the $S_k$, every $\pm 1$ will
carry {\it internal} indices which
indicate the type of quark and the color. These internal variables provide
additional degrees of freedom through weak interactions (flavor change), and
photon and gluon production. The latter two are special in that the
electromagnetic field allows quark/anti-quark pair annihilation, which is
equivalent to the introduction of finite life time lattice defects.
The gluons facilitate quark/anti-quark pair flavor change and pair
production, as well as quark color conversion.
Note in this that the sole emission of a gluon by a quark leads to the
creation of a quark/anti-quark pair at an empty pair of lattice sites.
The formal coupling constant in (11) is unity because the metric on the
four-manifold bounded by any three-torus in the lattice is Minkowskian,
i.e.\ $L(T^3)$ is metrically self-dual, and therefore $\kappa =\kappa^{-1}$.
Powerful Monte Carlo techniques developed for condensed matter physics on
frustrated lattices should be applicable to QCD physics.

\subsection{The Fine Structure and Other Coupling Constants}

The GUT energy scale $K=1/138$ in Planck units should define the QED coupling
constant since it reflects the emergence of individual particles, that can
travel along the $U(1)$ junctions, from the ground state $Q$. The measured
fine structure constant $\alpha_0=1/137.0359895$
agrees with $K$ to 0.7\%, where the discrepancy is due to the presence of a
non-trivial space-time topology. The manifold $Q$ supports
$\eta_Q=23^3+1$ dynamical degrees of freedom through the equation of motion,
where the 1 is for its time
evolution. For $L(T^3)$ one has $7^3+1$ dynamical degrees of freedom, but the
lattice has a four-fold degeneracy, yielding $\eta_L=e_{\rm M}^{-1}(7^3+1)=86$
(indeed an integer). This gives
$\alpha_K=K(\eta_Q-1)/(\eta_Q-\eta_L)=1/137.0359168$ for the
fine structure constant, corrected for those degrees of freedom
associated with the excited state character of the Universe, which do not
contribute to the ground state at any point in time. There is another
correction associated with the fact that the total system, ground state plus
excited state and the four homotopically distinct paths through the lattice,
has $\eta_{\rm t}=e_{\rm M}(\eta_Q-\eta_L)\eta_L$
effective degrees of freedom, but
quantization, fixed $\ell_{\rm Planck}$ and $m_{\rm Planck}$, removes two of
these to define a physical theory.
This gives a correction factor $f=(\eta_{\rm t}-2)/\eta_{\rm t}$.
One finds $\alpha =f\alpha_K=1/137.0359828$, with a relative accuracy of
$4.9\times 10^{-8}$, consistent with the experimental $1\sigma$ uncertainty in
$\alpha_0$ of $4.5\times 10^{-8}$. The numerical value of $\alpha$ then fixes
the unit of charge $e$. Note that the constancy of the speed of light is an
automatic consequence of the nuclearity of the electroweak sector $\Theta$[2],
and hence one does not have to correct for it.

The weak coupling constant follows from
$g^2_{\rm W}/4\pi =4/7\cdot 86/85\alpha_0$ because the ratio of these coupling
constants in the same electroweak sector must reflect the respective number
of degrees of freedom of the lattice ($e_{\rm M}=4$ paths between any two
three-tori, long range) and $T^3$ (multiplicity of seven, short range),
and is modified for the photon degree of freedom. With the standard definition
of the weak mixing angle ${\rm sin}^2\theta_{\rm W}=\zeta$, this yields
$\zeta =8^{-1}[7/4\cdot 85/86]=0.216$, where $\zeta$ should be viewed as
including all higher order quantum corrections.
One thus finds the masses for the vector bosons\hfill\break
$m_{\rm W}={\rm sin}^{-1}\theta_{\rm W}(\pi\alpha /G\surd 2)^{1/2}=80.18$ GeV
and
$m_{\rm Z}=2{\rm sin}^{-1}2\theta_{\rm W}(\pi\alpha /G\surd 2)^{1/2}=90.56$
GeV, consistent with experimental limits and the $7^{-3}$ intrinsic uncertainty
associated with the collapse of their rest mass wave functions.

It is easy to see that the measured value of the weak mixing angle is
${\rm sin}^2\theta^0_{\rm W}=(1+{\cal A}/3)\zeta =0.23246$, corrected for
the intrinsic accuracy ${\cal A}$ of $Q$ (see [2]). This correction is a result
of the fact that the vector bosons are represented in the ground state, but
derive their masses from the non-zero vacuum expectation value of the Higgs
field, which is a property of the excited state (this reasoning is entirely
topological). The value for ${\rm sin}^2\theta^0_{\rm W}$ is in poor agreement
with the overall mean experimental value of $0.23148\pm 0.00021$, but in good
agreement with the LEP average of $0.23196\pm 0.00028$.

Finally, the Fermi coupling constant $G=\surd 2 (g_{\rm W}/m_{\rm W})^2$ is
fixed through\hfill\break
$G^{-1/2}=-m_{\rm H}+\Sigma_i m_i=293.2$ GeV, i.e.\ by the sum of
all the particles represented in the ground state minus the
Higgs boson. The latter reflects spontaneous symmetry breaking, which is
associated with the excitation of the ground state[2]. This value
$G=1.163\times 10^{-5}$ GeV$^{-2}$ is consistent with the experimental limit of
$G_0=(1.16632\pm 0.00002)\times 10^{-5}$ GeV$^{-2}$, given the intrinsic
$7^{-3}$ uncertainty in the collapsed rest mass wave functions. In fact, since
some of the mass terms on the right hand side depend upon the value of $G_0$,
{\it a disagreement would have ruled out TD theory immediately}.

\section{A Spectral Observation of Planckian Topology through the Vacuum Energy}

In [2] it is shown that $L(T^3)$ possesses energy levels corresponding to
integer multiples $m$ of $\nu_0$
$$\nu_m =857.3588m\quad {\rm MHz},\eqno(12)$$
or $34.96698/m$ cm, where the numerical accuracy is limited by the Planck mass.
When these levels, are excited through matter degrees of freedom, they violate
T and CP invariance (e.g.\ the kaon system). The zero point energy, for
${\cal A}=(14/23)^3$, is given by
$$E_0={{14}\over{23}}{{(1+{\cal A}/3)m_{\rm Planck}}\over{23!14!}}\quad {\rm eV},\eqno(13)$$
and reflects the detailed structure of $Q$ through its numerical coefficients.
The factor $1/23!14!$ corrects for the total number of permutations of a
neutral field on $Q$. The $(14/23)$ denotes the splitting of the resulting
level in a $L(T^3)$ part and a $\Theta$ part, where the latter evaporates and
is related to $\Lambda$ (as discussed in Section 4)
and massive particles. The shift
induced by ${\cal A}/3$ signifies that the ground state $Q$ is three-fold
degenerate through the $Z_3$ of $\Theta$ (the $1/3$), while at the same time
only an accuracy of $14^{-3}$ on $P$ is required to convey the information
present on $Q$, with an accuracy of $23^{-3}$, to the individual $U(1)$
junctions on $L(T^3)$.
The excited states correspond to vibrations of the junctions in $L(T^3)$, and
are therefore homeomorphic but not diffeomorphic. Energy considerations then
imply that photons of the appropriate wavelength can also interact with these
levels on any $U(1)$ lattice junction since they are massless.

A photon of the required energy will then
have a partner whose frequency $\nu_m'$ obeys $\nu_m'=\nu_0/m$. This is a
fixed point pairing under $E_0$ of $k^\mu k_\mu =0$ for the four vector
$(2\pi\nu_0,k_i)$. That is, for every
$\nu_m$ one should find a ghost image at frequency $\nu_m/m^2$ with a
wave vector differing only in scale. The duality
implies that destruction of either partner leads to destruction of the pair.
Furthermore, the pairing is not induced by photons with energies below $E_0$
since the suggested effect is quantized on the natural numbers.
Clearly, the distinguishing character of this signal resides in the fact that
its strength is always equal to that of the ``source'' spectrum at the high
frequency. At the frequency $\nu_0$ one can observe no effect, since detection
of the one photon implies removal of the partner. Furthermore, the high
frequency spectral features (so not their duals) are also special in that their
line width is always equal to $7^{-3}\nu_0=2.4996$ MHz.

Finally, the zero-point $E_0$, as well as being the characteristic energy of
the lattice vibration, is a vacuum energy which behaves like a cosmological
constant. Its contribution is $\sim 10^{-58}$ GeV$^4$, and hence it is
negligible. To summarize the above sections on $\Lambda$, the TD of space-time
leads to quantities which behave like a cosmological constant through black
hole singularities and $L(T^3)$. Indeed, in the context of TD there are two
manifestations of $\Lambda$, but one underlying topological structure.

\section{Discussion}

In conclusion, an additional set of diagnostics has been derived from the TD
approach. Some of these are found to be in excellent agreement with current
experiments or observations, others will have to await more sophisticated
measuring techniques. A straightforward test is the spectral signature,
which has no counterpart in the Standard Model, and whose observation would
confirm, barring a major fluke, the existence of $Q$ as the ground state.

\section*{Acknowledgements}
The author is indebted to J.A.A.~Berendse-Vogels,
G.~van Naeltwijck van Diosne, and M.C.~Spaans sr.\ for valuable assistance.
This work was supported by NASA through Hubble Fellowship grant HF-01101.01-97A
awarded by the Space Telescope Science Institute, which is operated by the
Association of Universities for Research in Astronomy, Inc., for NASA under
contract NAS 5-26555.

\end{document}